\documentstyle[psfig,PASJadd]{PASJ95kai}
\newcommand{\lett}{}

\newcommand{\rdate}{}
\newcommand{\adate}{}

\begin{document}

\setcounter{page}{1}

\title{Study of the Long-Term X-Ray Variability of a Possible Quasar
 RX~J0957.9+6903 with ASCA}

\author{Yu-ichiro {\sc Ezoe}$^1$, Naoko {\sc Iyomoto}$^2$, 
and Kazuo {\sc Makishima}$^1$
\\[12pt]
$^1${\it Department of Physics, University of Tokyo, 
         7-3-1 Hongo, Bunkyo-ku, Tokyo 113-0033}\\
{\it E-mail(YE): ezoe@amalthea.phys.s.u-tokyo.ac.jp }\\
$^2${\it The Institute of Space and Astronautical Science,
        3-1-1 Yoshinodai, Sagamihara, Kanagawa 229-8510}\\
}

\abst{

Long-term variability and spectral properties of
a possible quasar, RX~J0957.9+6903, 
 were studied, utilizing 16 ASCA observations spanning 5.5 years.
The average 0.7--10 keV spectrum of RX~J0957.9+6903
is well represented by a power-law continuum of photon index
1.58 $\pm$ 0.03, and an absorption column of
$\sim$ 1$\times{\rm 10}^{21}$ ${\rm cm}^{-2}$.
The 2--10 keV flux of RX~J0957.9+6903 varied by a factor of four
over the period of six years, around a mean of
$\sim$ 8.8$\times {\rm 10}^{-12}$ erg ${\rm s}^{-1}$ ${\rm cm}^{-2}$.
Peak to peak variability within each observation was less
than 25\% on $\sim$ 1 day time scale.
These properties support the classification of
RX~J0957.9+6903 as a quasar.
The power spectrum density (PSD) was estimated in a "forward" manner,
over a frequency range of $10^{-8.2}$--$10^{-4.3}$ Hz,
by utilizing the structure function method 
and Monte-Carlo simulation assuming a broken power-law type PSD.
Then, the break frequency $f_{\rm b}$ of the PSD of
RX~J0957.9+6903 has been
constrained as 1/$f_{\rm b}$ = $1600^{+\infty}_{-1100}$ days,
 and the logarithmic slope of the high-frequency region of the PSD as 
$\alpha$ = $-$1.55 $\pm$ 0.2 .
A comparison of the estimated PSDs is made between
RX~J0957.9+6903 and the M81 nucleus, observed in the
same field of view.
}

\kword{Galaxies: active
-- Galaxies: individual (RX~J0957.9+6903)
-- X-rays: galaxies
-- Methods: structure function
}
\maketitle


\section{Introduction}

Emission from an active galactic nucleus (AGN) is known to exhibit
apparently random variability, over a wide wavelength range from 
radio to X-rays and $\gamma$-rays (e.g., 
Krolik et al. 1991; Edelson et al. 1996).
Although the exact origin of such a variability is still unclear, 
it has been employed practically as a measure of
the mass and size of the central black hole (BH) residing in the AGN. 
For example, Wandel \& Mushotzky (1986)
pointed out a strong correlation between
the mass of the BHs in the AGNs estimated from optical emission lines,
and the timescale of their X-ray variability
in terms of the intensity  doubling time.
Hayashida et al. (1998) showed that the more luminous AGNs, which may
have more massive central BHs, have longer timescale of variability;
they utilized power spectrum density (PSD),
which is mathematically more exact 
than the doubling time but suffers from the window function
(the Fourier transform of the observational sampling) convolved with
the true power spectrum of the source. 
These results indicate that the intrinsic X-ray variability 
timescale of AGN is an important parameter,
which roughly represents the mass of the central BH.

Among AGNs, high-luminosity ones including quasars (QSOs) have such
a long timescale of variability, e.g. a few years, that   
the analysis of their variability timescale needs long observations. 
Such long observations in X-rays are generally limited to
the all-sky monitoring of bright sources,  
including a limited number of QSOs
(e.g., $>$ 30 mCrab for RXTE, where "1 mCrab" flux is 
$\sim$ 3 $\times$ $10^{-11}$ erg s$^{-1}$ cm$^{-2}$ 
in the 2--10 keV range).

RX~J0957.9+6903, a possible QSO, 
has been observed as many as 16 times over six years
from 1993 to 1999 with ASCA, in which the total exposure time reaches
620 ksec. This is because the object lies close to
the spiral galaxy M81, and was always within the same field-of-view
(FOV) in the repeated ASCA observations of SN1993J. 
This object was first detected in the Einstein observation of M81,
and was designated M81 X-9 (Fabbiano 1988; Elvis et al. 1992),
although it is not associated with M81.
Subsequently it was observed with  EXOSAT (Giommi et al. 1991) and 
ROSAT, and renamed RX~J0957.9+6903 (Bade et al. 1998).
It was identified optically with a faint point-like object
having an extreme blue continuum (Bade et al. 1998)
in the wavelength range between 3400 $\AA$ and 5400 $\AA$.
These optical properties suggest that RX~J0957.9+6903 is a QSO,
even though its cosmological redshift is still unknown.
On the other hand, its association with
a possible supernovae remnant (SNR), 
 at a separation of $\sim$ 10$^{\prime\prime}$, was proposed 
from an H$_\alpha$ observation (Miller et al. 1995).

In the present paper, we first study the 0.7--10 keV X-ray spectrum of
RX~J0957.9+6903 and variability in section 3, and show them to be
consistent with those of a QSO. 
We then analyze the long-term light curve of RX~J0957.9+6903
more quantitatively in section 4,
and estimate its PSD utilizing the new method, developed by
Iyomoto et al. (2000) and Iyomoto (1999) to study
the long-term variation of the M81 nucleus.
Because RX~J0957.9+6903 and the M81 nucleus have always been
in the same ASCA FOV and observed under the same sampling 
window, we can directly compare the estimated PSD of these two objects, 
one being a promising QSO candidate while
the other a typical low-luminosity AGN (LLAGN).

\section{Observation}

We observed the M81 region 16 times with ASCA as summarized in table 1.
It is among the objects most frequently observed with ASCA.
We analyze only data from the GIS (Gas Imaging Spectrometer; 
 Ohashi et al. 1996; Makishima et al. 1996)
with FOV of $50^{\prime}$ diameter,
because our target, RX~J0957.9+6903,
was always outside the SIS (Solid-State Imaging Spectrometer) FOV in
even the 4-CCD mode ($22^{\prime}\times22^{\prime}$ FOV).
The GIS data were acquired in PH normal mode throughout.

For the GIS event selection,
we required the geomagnetic cut-off rigidity
to be $>$ 8 GeV ${\rm c}^{-1}$ and
the target elevation angle to be
 $>$ $5^{\circ}$ above the horizon.
These criteria left us with 620 ksec
of good GIS data. RX~J0957.9+6903 was always detected, at
a mean 0.7--10 keV counting rate of
6 $\times 10^{-2} $ c ${\rm s}^{-1}$
for each GIS detector (GIS2 or GIS3).

In figure 1 we show an example of the GIS images
of the M81 region, after
subtracting the non X-ray background.
There were always two discrete X-ray sources in the GIS FOV;
the brighter one (to the right in figure 1) is the emission complex
from M81, while the fainter one (to the left in figure 1)
is RX~J0957.9+6903.
The latter was detected at a J2000 position 
($9^{\rm h}57^{\rm m}43^{\rm s},
+69^{\rm \circ}04^{\rm \prime}03^{\rm \prime\prime}$),
which is consistent with the J2000 optical position 
($9^{\rm h}57^{\rm m}51^{\rm s}\hspace{-4pt}.\hspace{2pt}0,
+69^{\rm \circ}03^{\rm \prime}30^{\rm \prime\prime}$; Bade .et al. 1998)
within the position accuracy of ASCA 
($\sim 1^{\prime}$; Tanaka et al. 1994).
The M81 complex is elongated downward and upward because
it consists of the nucleus (X-5) and three additional sources,
SN 1993J, X-6, and X-4;  these four sources are
not resolved clearly 
from one another in the GIS image of figure 1.
The results on SN 1993J have been reported by
Kohmura (1994) and Kohmura et al. (1994).
The results on the M81 nucleus of the 1st to 10th observations
have been published as Ishisaki et al. (1996).
The details of timing analysis of the M81 nucleus of
all the 16 observations are published as Iyomoto et al. (2000).

\section{X-Ray Properties of RX J0957.9+6903 }

\subsection{Spectrum}

For each observation, we accumulated photons within $3^{\prime}$ 
of the X-ray centroid of RX~J0957.9+6903
in the image and made spectra.
The contamination from the M81 complex, estimated
with the point spread function of the XRT
(X-ray telescope), is less
than $\sim$ 1\% of the total counts of RX~J0957.9+6903 
in this region, and can be negligible.
Therefore, the background was taken from the same detector
region in blank skies and subtracted,
the count rate of which was an order of magnitude
less than that of RX~J0957.9+6903.
The XRT+GIS response files were calculated separately 
for the two instruments (GIS2,3).
We then added the GIS2 and GIS3 spectra into a
single spectrum for each observation.

The obtained X-ray spectra of RX~J0957.9+6903 are
relatively featureless and all alike.
In order to obtain a rough idea of the spectral variation,
we fitted the 0.7--10 keV GIS (GIS2+GIS3) spectrum of individual
observations with an absorbed power-law model.
The absorption column density and photon index 
were taken as free parameters except the 16th observation.
These 15 fittings were acceptable with the reduced $\chi^2$ $\sim$ 0.7.
The average absorption column density of these 15 spectral fittings 
was obtained as $\sim 9.5 \times 10^{20}$
 ${\rm cm}^{-2}$,
which is higher than the Galactic line-of-sight column
density of $4.06 \times 10^{20}$ ${\rm cm}^{-2}$.
The average photon index turned out to be $\sim 1.6$
except the 13th observation,
when the photon index was 2.2.
For the spectrum of the 16th observation,
the column density became lower
than the Galactic value and therefore was fixed
at that Galactic value, which yielded an acceptable fit.
The 2--10 keV fluxes and photon indices derived in this way 
from individual observations are 
shown in figure 2a and figure 3, respectively,
as a function of time.

Because the spectral variation is thus negligible except
the 13th and 16th observations, we summed up all
the GIS (GIS2+GIS3) data except the 13th and 16th
into a single spectrum.
We calculated the average XRT+GIS responses
as averages of those for individual observations weighted by
respective exposure times.
We also calculated the average background.
Using these response and background,
we fitted the summed GIS spectrum
in the 0.7--10 keV range with an absorbed power-law model.
This model was acceptable with the reduced $\chi^2$ of 0.92.
The obtained best-fit model is shown in figure 4 and the best-fit
parameters are given in table 2.

As shown in figure 4b, a similarly good fit
(reduced $\chi^2$ of 0.75)
to the average spectrum has been obtained by replacing
the power-law model with a bremsstrahlung model of temperature
$kT = 12.0\pm0.8$ keV, with the absorbing column density
fixed at the Galactic line-of-sight value.
In table 2, we list these fit parameters as well.

\subsection{Light curves}

In figure 2a we present the background-subtracted
long-term light curve of RX~J0957.9+6903,
in terms of the 2--10 keV flux as derived in section  3.1.
Each data point corresponds to each of the 16 observations.
Thus, the long-term variation amounts to a factor of
$\sim$4 (peak-to-peak), around an average flux of
8.8$\times10^{-12}$ erg cm$^{-2}$ s$^{-1}$.

To investigate the short-term variability,
we subdivide each data point
in the light curve into finer bins of  1/4 day width.
Then, the peak-to-peak variation in each observation was
found to be less than $\sim$ 15\% except the 13th observation,
when it was 25\%.
Figure 2b shows the short-term light curve
of the 13th and 3rd observations,  
the latter being the longest one.
With this bin size, all the 16 observations
constitute a 51-bin light curve.
Typical 1$\sigma$ error due to photon counting
statistics is $\sim$ 4\% of the average flux, whereas 
root-mean-square (RMS) of overall variation through
the entire light curve is much larger, 30\% of the mean.
This is important for our structure function analysis
as described in section 4.1. 

\subsection{Comparison with the past observations}

The flux of RX~J0957.9+6903 measured in past observations
is summarized in table 3.
The results of the Einstein and EXOSAT data is quoted from Fabbiano (1988)
and Giommi et al. 1991, respectively.
The EXOSAT flux is converted from the countrate by utilizing the conversion
factor in Fig 2 of Giommi et al. 1991.
We extended the best fit power-law model of ASCA shown in table 2,
to estimate the 0.2--4.0 , 0.05--2.0 and 0.5--2.0 keV fluxes.
Below we explain the ROSAT data.

We newly analyzed the three ROSAT PSPC archive data in table 3,
which cover epochs between the Einstein and ASCA observations.
The background-subtracted PSPC spectra of RX~J0957.9+6903
are show in figure 5.
The background spectra were taken from the blank field
in the same FOV of each observation.
These spectra were represented by an absorbed power-law
model with the photon index $\sim 2$ and the absorption
$\sim$ 2 $\times 10^{21}$ ${\rm cm}^{-2}$.
The reduced $\chi^2$ were 9/13, 5/10, and 21/16
for the 1991 May, October, and 1992 September observations,
respectively. 

Compared with the estimated ASCA fluxes,
the Einstein, EXOSAT  and ROSAT values are lower
by a factor of $\sim 1.5$.
Considering, however, uncertainties of
the cross-calibration between different instruments operating 
in different bands, and the RMS $=30\%$ of 
the ASCA light curve,
these fluxes measured previously can be
thought to be within the variation range of RX~J0957.9+6903.

\subsection{The nature of RX~J0957.9+6903}

We have for the first time measured
an accurate 0.7--10 keV spectrum of
RX~J0957.9+6903, and derived short/long-term light curves. 
The obtained results, i.e., a power-law type spectrum
with $\Gamma\sim 1.6$,
small short-term variability and
a significant long-term variability, 
are consistent with the past suggestion 
of RX~J0957.9+6903 being a QSO
(Fabbiano 1988; Ishisaki et al. 1996; Bade et al. 1998).
These hard spectra and the random variability over $\sim$ 6 years
rule out the association of this object 
with an SNR (Miller et al. 1995).
The object can not be a Seyfert galaxy since
the optical counterpart is point-like (Bade et al. 1998).
Although the featureless blue optical continuum is
consistent with those of Blazers,
the lack of a radio source within  $\sim$ 1$^{\prime}$
of RX~J0957.9+6903 (from NED) rules out its Blazer interpretation.

 Although RX~J0957.9+6903 has a high Galactic latitude
($\sim$ 40$^{\circ}$), there still remains a small possibility
that it is a Galactic object.
The successful fit to the X-ray spectrum with a high-temperature
Bremsstrahlung (figure 4b) leaves room for either a cataclysmic
variable (CV) or a low-mass X-ray binary (LMXB);
the former exhibits a genuine high-temperature thermal X-ray spectrum
(e.g., Ishida 1991), while the latter an optically-thick
spectrum which can be approximated
by a mildly absorbed ($N_{\rm H}$ $\sim$
1$\times10^{21}$ cm$^{-2}$) high-temperature Bremsstrahlung
(Makishima et al. 1989).
The featureless blue optical continuum of
RX~J0957.9+6903 (Bade et al. 1998)
is inconsistent with the CV interpretation, while consistent with
the LMXB scenario where the optical emission mainly
arises from reprocessed X-rays.
However, in addition to the small number of high-latitude
Galactic X-ray sources,
its X-ray to optical flux ratio becomes
log $[f_{\rm x}/f_{\rm B}] \sim 0.9$ where $f_{\rm B}$ is
the optical blue flux obtained
from ${\rm m}_{\rm B}$=18.6 (Bade et al. 1998). This is low for an
LMXB which would show log $[f_{\rm X}/f_{\rm B}] \sim  2$ -- 3
; e.g.,  Ritter 1990, while consistent with that of QSO
( log $[f_{\rm X}/f_{\rm B}] \sim$ $-1$ -- +1;
e.g., Tananbaum et al. 1979).

From these considerations, we conclude that
RX J0957.9+6903 is a radio-quiet QSO. 
If, for example, the object lies at a redshift of 1.0,
the 2--10 keV luminosity
becomes 3$\times10^{46}$ erg ${\rm s}^{-1}$,
assuming the Hubble constant
to be 70 km s$^{-1}$ Mpc$^{-1}$ and a flat universe
without cosmological
constant. This is reasonable as a QSO luminosity.


\section{Structure Function Analysis}

We obtained a 51-bin light curve  with the 1/4-day bin width.
Because of the sparseness of this light curve, 
it is difficult to determine the PSD of RX~J0957.9+6903 directly
by Fourier transformation.
Therefore we utilized the "forward method" analysis incorporating
structure function as developed by Iyomoto (1999)
and Iyomoto et al. (2000). Below we briefly explain this method.
Further details of this method are given by Iyomoto et al. (2000).

\subsection{Structure function}
For a time series of luminosity $l_m$, the first-order
structure function (SF), $S(\tau_k)$, 
is given as

\begin{equation}
\begin{array}{rcl}
S(\tau_k) &=&\displaystyle
\frac{1}{N(\tau_k)}\sum_{m=0}^{N-1-n}(l_m - l_{n+m})^2 \\
&&\\
\end{array}
\label{eqn:SFdef}
\end{equation}
where $l_n$ is the luminosity at time $t_n$ and $\tau_k$ is the time lag
between $t_m$ and $t_{n+m}$. The sum is taken over those pairs
which have the same time lag $\tau_k$, and $N(\tau_k)$ is
the number of such pairs.
The general definition of SF is given by Simonetti et al. (1985).
In the case of equally sampled data,
SF and PSD are mathematically equivalent.

Figure 6a shows the SF of RX~J0957.9+6903
calculated from the 51-bin
light curve using equation (\ref{eqn:SFdef}).
The statistical error of the light curve contributes little, 
because the RMS variation of the light curve is larger 
by an order of magnitude as described in section  3.2;  
the contribution of the Poisson noise to the SF is calculated to be
[(Poisson error)/(RMS)]$^2 =(4/30)^2 \sim $ 2\%.
In figure 6a we can see that the various sampling intervals give
the wide-band SF. The SF is smooth at short time lags because
there are many pairs with such small time lags,
or $N(\tau_k)$ is large.
The SFs at  large time lags scatter very much
because of small values of $N(\tau_k)$.
To suppress the scatter of SFs at large time lags,
we have binned the SFs into 22 appropriate intervals
as shown in figure 6b.
Thus, the SF keeps increasing rather monotonically
to longer time lags,
indicating that the typical time scale of variation is long.
Even using the binned SF, however, it is difficult to convert the SF
into PSD because of the sampling window effect of
the 16 observations.
Therefore we utilize the forward method analysis.

\subsection{Forward method simulation}
As the forward method analysis,
we begin with producing many Monte-Carlo light curves
assuming a model PSD and the actual observation sampling window.
We next convert each simulated light curve into a fake SF and then
compare the ensemble average of those fake SFs with the SF
derived from the actual data.

As the PSD model of RX~J0957.9+6903,
we assume a broken-power-law shape as
\begin{equation}
P(f) = \left\{
\begin{array}{ll}
P_0 & ( f<f_{\rm b} ) \\
P_0(f/f_{\rm b})^{\alpha} & ( f>f_{\rm b} ) 
\end{array}
\right.
\label{eqn:moPSDdef}
\end{equation}
where $f$ denotes frequency, $P_0$ is a constant, $f_{\rm b}$ is a
characteristic frequency called "break frequency", and $\alpha$ is a
slope index. It is an empirical PSD of AGNs
(e.g., Pounds \& McHardy 1988; Edelson \& Nandra 1999).
We regard $f_{\rm b}$ and $\alpha$ as free parameters.
The trial values of
1/$f_{\rm b}$ are $10^{1.0}, 10^{1.2}, 10^{1.4}$ ... and $10^{3.4}$ days,
and those of $\alpha$ are $-1.20$, $-1.25$, $-1.30$ ... and $-2.40$.
For each pair of $f_{\rm b}$ and $\alpha$,
we generated 1000 fake light curves
via Monte-Carlo method, by giving random phases
to the Fourier components.
Each fake light curve was subjected to
the sampling window for RX~J0957.9+6903,
to make a simulated 51-point sparse light curve.
Then we calculated an SF from each fake 51-point light curve. 

Figure 6c shows the ensemble average
of the 1000 simulated SFs, for $\alpha = -1.55$ and
$f_{\rm b} = 10, 100$, and $1000$ days, 
without applying the sampling window.
Comparing figure 6b with figure 6c, we can again estimate
that the variation timescale of RX~J0957.9+6903 is relatively long.
Figure 6d shows the ensemble average of the 1000 fake SFs
which were calculated after each simulated light curves was
multiplied with the actual sampling window; 
it was subjected to the same binning as figure 6b, so that
the simulation can be directly compared to the actual SF.

We compared the observed and simulated SFs
by utilizing the $\chi^2$ technique.
The $\chi^2$ is defined as

\begin{equation}
\chi^2 = \sum_{\tau_k}{ \frac{(S_{\tau_k}^{\rm Obs} -
\overline{S_{\tau_k}^{\rm Fake}} )}{\sigma^{\rm Fake}_{\tau_k}} }
\label{eqn:chi2def}
\end{equation}
where $S_{\tau_k}^{\rm Obs}$ and
$\overline{S_{\tau_k}^{\rm Fake}}$ are
the observed and the average fake SFs, while
$\sigma^{\rm Fake}_{\tau_k}$
denotes RMS of the 1000 fake SFs at time lag $\tau_k$.
We regard this $\chi^2$ as a measure of goodness of
the employed PSD model.
Among various PSDs which have different values of
$f_{\rm b}$ and $\alpha$,
we have obtained the best PSD for $1/f_{\rm b}$ = $10^{3.2}$ days and
$\alpha$ = $-1.55$ with $\chi^2/\nu =$ 0.84.
We can regard the PSD with these parameters to be the best
as long as assuming the model PSD of equation (\ref{eqn:moPSDdef}).

\subsection{Uncertainties in the power spectrum density}

We again utilized Monte-Calro simulation to evaluate
uncertainties in the PSD estimation.
We generated another 1000 fake light curves using the best
parameters, $f_{\rm b}$ and $\alpha$, obtained in section 4.2.
The random numbers of these simulated light curves
are different from each other, and from those used in section  4.2.
For each of the newly generated fake SFs,
we determine the optimum $f_{\rm b}$ and $\alpha$
in the same way as for the RX~J0957.9+6903 data,
and derive the $\chi^2$ value,
$\chi_{\rm min}^2$. We derived also $\chi^2$, $\chi^2_{\rm true}$, 
on each simulation at the best-fit parameters of
$1/f_{\rm b}=10^{3.2}$ days
and $\alpha=-1.55$, obtained in section 4.2. We then calculate

\begin{equation}
\Delta\chi^2 \equiv \chi^2_{\rm true} - \chi_{\rm min}^2
\end{equation}
for each light curve.
Figure 7a shows the integrated distribution of
 $\Delta\chi^2$ from the latter set of 1000 simulations.
We determine the 90\% and 99\% confidence regions from this distribution.
In figure 8a, we show the obtained confidence regions
on the 2-dimension plane of $f_{\rm b}$ and $\alpha$.
The best parameters with 90\% confidence errors thus become 
1/$f_{\rm b}$ = $1600^{+\infty}_{-1100}$ days and
 $\alpha$ = $-1.55$ $\pm$ 0.2. Similarly in figure 8b,
we present an overlay of various PSDs corresponding to
the 90\% confidence region of figure 8a.
We can see a possible break in the PSD.

Utilizing the $\chi^2_{\rm true}$ distribution
of the 1000 simulations,
we can also estimate how well the assumed PSD expresses
the RX~J0957.9+6903 PSD.
Figure 7b shows the distribution of $\chi^2_{\rm true}$,
where an arrow indicates the $\chi^2$ value we have obtained
from the SF of RX~J0957.9+6903 in section 4.2.
The $\chi^2$ value of RX~J0957.9+6903 is almost at the peak
of this distribution,
which justifies our assumption of  equation (\ref{eqn:moPSDdef}).

\subsection{Time scale of variation of RX~J0957.9+6903}
We have estimated the break frequency and slope index of
RX~J0957.9+6903 on the assumption of equation (\ref{eqn:moPSDdef})
which describes typical intensity variations in AGNs.
Seyfert galaxies exhibit $f_{\rm b} = 10^{-4}$ $\sim$ $10^{-6}$ Hz
and $\alpha$ = $-1$ $\sim$ $-2$ (e.g., Lawrence et al. 1987
for NGC 4051; McHardy, Czerny 1987 for NGC 5506;
Hayashida et al. 1998 for NGC 4051 and
MCG 6--30--15). From Galactic black hole candidates,
we typically observe
$f_{\rm b} \sim$ 0.1 Hz and $\alpha =$ $-1$ $\sim$ $-2$
(e.g., Makishima 1988).
Our results for RX~J0957.9+6903 are
$f_{\rm b} \sim 7.3 \times 10^{-9}$ Hz
and $\alpha \sim -1.55$.
This $\alpha$ value is consistent with those of typical AGNs.
Furthermore, the variation time scale ($1/f_{\rm b}$)
of RX~J0957.9+6903 is about 3 orders of magnitude longer
than those of rapidly varying Seyferts.
This long term variation resembles those of 3C273,
a representative QSO (Hayashida et al. 1998).

Although the physical mechanism accounting for equation
(\ref{eqn:moPSDdef}) is not yet understood,
lower values of $f_{\rm b}$ are generally thought to imply
higher central BH masses (e.g., Hayashida et al. 1998).
If we further assume that the BH mass $M_{\rm B}$ is
inversely proportional to $f_{\rm b}$, it may be written as
\begin{equation}
M_{\rm BH} = 10 \times 0.1 [{\rm Hz}] / f_{\rm b}\hspace*{0.5em}M_{\odot}
\label{eqn:BHMass}
\end{equation}
where 0.1 Hz and 10 $M_{\odot}$ are the parameters
of Cyg X-1 as a standard.
Then our results on RX~J0957.9+6903 imply $M_{\rm B} > 10^{8} M_{\odot}$.
This value is in a typical range of mass of QSOs estimated
from their luminosities, which are assumed to be close
to the Eddington limit
(e.g., Lawson \& Turner 1997, Reeves et al. 1997).

\subsection{Comparison with the M81 nucleus}
Utilizing the same structure function method, 
Iyomoto et al. (2000) have analyzed the time variability of
the M81 nucleus, an LLAGN,
detected in the same FOV as RX~J0957.9+6903
over these 6 years.
Because the window function for RX~J0957.9+6903 is
the same as that for the M81 nucleus, we may directly compare
the time scales of their variations.
Iyomoto et al. (2000) have found the inverse of the break frequency of the M81 nucleus
to be larger than 800 days (90\% lower limit),
which is close to that of RX~J0957.9+6903,
$1600^{+\infty}_{-1100}$ days.
Therefore the M81 nucleus (LLAGN) may have a central BH as heavy as
that of RX~J0957.9+6903 (QSO). This in turn suggests that
the central BH mass of the M81 nucleus, and generally
of other LLAGNs, is higher than those of
the fast-varying Seyfert nuclei,
and their very low luminosities are
a result of extremely small accretion rates as argued, e.g., by
Iyomoto et al. (1996), Ishisaki et al (1996),
and Iyomoto et al. (1998).

\section{Summary}

RX J0957.9+6903 has been observed 16 times with ASCA over 5.5 years. 
The average 0.7--10 keV GIS spectrum 
is represented by a power-law continuum of photon index  
1.58 $\pm$ 0.03, and an absorption column of $\sim$ 
1$\times{\rm 10}^{21}$ ${\rm cm}^{-2}$.
Spectral variation was insignificant
except the 13th and 16th observation.
The 2--10 keV flux varied by a factor of four through 6 years, and the 
short-term variation was 25 \% or less on a time scale of about
one day. These results, together with the optical observation, 
support the interpretation of RX~J0957.9+6903
as a QSO (Fabbiano 1988; Bade et al. 1998).

Through a forward-method analysis utilizing structure function, 
the PSD of RX J0957.9+6903 has been estimated
to have a very low break
frequency $f_{\rm b}$, $1/f_{\rm b}$ =
$1600^{+\infty}_{-1100}$ days, and the slope index
$\alpha$ = $-1.55 \pm$ 0.2.
From the mass-timescale relation for various BHs,
the central BH mass of RX~J0957.9+6903 is estimated
to be $> 10^{8}$ $M_{\odot}$.
The comparison of the PSDs between RX~J0957.9+6903 (QSO)
and the M81 nucleus (LLAGN),
estimated in the same way under the identical sampling window,
indicates that the two objects have similar time scales of variation.
This further supports that the central BH mass is not much different
between QSOs and LLAGNs.

\bigskip


\clearpage
\section*{Reference}
\vspace*{1em}

\small

\re
Bade N., Engels D., Voges W., Beckmann V., Boller Th., Cordis L., Dahlem  M., Englhauser J.,
Molthagen K., Nass P., Studt J., Reimers D. 1998, A\&A 127, 145 

\re
Edelson R. A., Alexander T., Crenshaw D., M., Kaspi S., Malkan M. A., Peterson B. M.,
 Warwick R. S., Clavel, J. et al. 1996, ApJ 470, 364

\re
Edelson R., Nandra K. 1999, ApJ 514, 682

\re
Elvis M., Plummer D., Schachter J., Fabbiano G. 1992, APJS 80, 257

\re
Fabbiano G. 1988, ApJ 325, 544

\re
Giommi P., Tagliaferri G., Beuermann K., Branduardi-Raymont G., Brissenden R.,
Graser U., Mason K., Mittaz J. et al 1991, ApJ 378, 77

\re
Hayashida K., Miyamoto S., Kitamoto S., Negor H., Inoue H. 1988, ApJ 500, 642

\re
Ishisaki Y., Makishima K., Iyomoto N. 1996, ApJ 48, 237

\re
Ishida. 1991, PhD Thesis, The University of Tokyo

\re
Iyomoto N., Makishima K., Fukazawa Y., Tashiro M., Ishisaki Y., Nakai N.,
Taniguchi Y. 1996, PASJ 48, 231

\re
Iyomoto N., Makishima K., Tashiro M., Inoue S., Kaneda H., Matsumoto Y.,
Mizuno T. 1998, APJ 503L, 31

\re
Iyomoto N. 1999, PhD Thesis, The University of Tokyo

\re
Iyomoto N., Makishima K. 2000, MNRAS submitted

\re
Kohmura Y. 1994, PhD Thesis, The University of Tokyo

\re
Kohmura Y. Inoue Y., Aoki T., Ichida M., Itoh M., Kotani T.,
Tanaka Y., Ishisaki Y. et al. 1994, PASJ 46, L157

\re
Krolik J. H., Horne K., Kallman T. R., Malkan M. A., Edelson R.A., Kriss G. A. 1991, ApJ 371, 541

\re
Lawrence A., Watson M.G., Pounds K.A., Elvis M. 1987, Nature 325, 694L

\re
Lawson A.J., Turner M.J.L. 1997, MNRAS 288, 920L

\re
Makishima K. 1988, in Physics of Neutron Stars and Black Holes, ed Y. Tanaka (Universal Academy Press, Tokyo) p175  

\re
Makishima K., Ohashi T., Hayashida K., Inoue H., Koyama K., Takano S., Tanaka Y., Yoshida A. et al. 1989, PASJ 41, 697

\re
Makishima K., Tashiro M., Ebisawa K., Ezawa H., Fukazawa Y.,
Gunji  S., Hirayama M., Idesawa E., et al. 1996, PASJ 48, 171
\re
McHardy I., Czerny B. 1987, Nature 325, 696M

\re
Miller B.W. 1995, ApJ 446, 75L

\re
Ohashi T., Ebisawa K., Fukazawa Y., Hiyoshi K., Horii M.,
Ikebe Y., Ikeda H., Inoue H., et al. 1996, PASJ 48, 157

\re
Pounds K.A., McHardy I.M. 1988 in Phisics of neutron stars and black holes, ed Y. Tanaka (Universal Academy Press, Tokyo) p.285

\re
Reeves J.N., Turner M.J.L., Ohashi T., Kii T. 1997, MRNAS 292, 468

\re
Ritter H. 1990, A\&AS 85, 1179

\re
Simonetti J.H., Cordes J.M., Heeschen D.S. 1985, ApJ 296, 46

\re
Tananbaum H., Avni Y., Branduardi G., Elvis M., Fabbiano G., Feigelson E., Giacconi R., Henry J.P., Pye J.P., Soltan A., Zamorani G. 1979, ApJ 234L, 9

\re
Tanaka Y., Inoue H., Holt S.S. 1994, PASJ 46L, 37

\re
Wandel A.,\& Mushotzky R.F. 1986, ApJ 306L, 61

\newpage

\bigskip

%
%
\onecolumn
\large
\vspace*{-10em}

\begin{table*}[t]
\begin{center}
Table 1. Log of ASCA observations of RX~J0957.9+6903.\\
\vspace*{1em}
\begin{tabular}{rccc}
\hline\hline\\
ID & 	\multicolumn{2}{c}{Date}            & GIS \\
\cline{2-3}
   & ${\rm Start}^{\ast}$  &${\rm End}^{\ast}$
&${\rm Exposure}^{\dagger}$ \\ \hline 
1  & 93/04/05 06:00 	   & 	04/06   00:53		&32745	\\
2  & 93/04/07 05:55        & 	04/07   22:11		&32065	\\
3  & 93/04/16 22:40        &    04/18   17:06		&88853 \\
4  & 93/04/25 14:15   	   & 	04/25   22:05		&12466  \\
5  & 93/05/01 22:17 	   & 	05/02   14:11		&24145 \\
6  & 93/05/18 20:48        & 	05/19   20:31		&40430  \\
7  & 93/10/24 14:33        & 	10/25   14:21		&43548  \\
8  & 94/04/01 05:08        & 	04/02   03:21		&38087  \\
9  & 94/10/21 04:25	   & 	10/22   07:41		&48263  \\
10 & 95/04/01 18:50	   & 	04/02   23:11		&21428  \\
11 & 95/10/24 18:40 	   & 	10/25   22:51		&36443  \\
12 & 96/04/16 05:34        & 	04/17   09:47		&48449  \\
13 & 96/10/27 04:13        & 	10/28   05:25		&32858  \\
14 & 97/05/08 12:24        & 	05/09   16:20		&50197  \\
15 & 98/04/10 06:18        & 	04/11   10:26		&48872  \\
16 & 98/10/20 19:54        & 	10/21   23:11		&23124  \\
\hline
\end{tabular}
\end{center}
\vspace*{1em}\par\noindent
\hspace*{4em}$\ast$ The start and the end time of the observation, in year/month/day hour:minitues and\\
\hspace{5em}month/day hour:minitues, respectively.
\par\noindent
\hspace*{4em}$\dagger$ Exposure time in seconds after the data screening.
\end{table*}

%
%

\vspace*{10em}

\begin{table*}[h]
\begin{center}
Table 2. Model fits to the average 0.7-10 keV spectrum of RX~J0957.9+6903$^{\ast}$.\\
\vspace*{1em}
\begin{tabular}{ccc}
\hline\hline
                                  & Power-law &  Bremsstrahlung\\ \hline
Photon Index or $kT$ (keV)        & 1.58 (1.55-1.61)   & 12.0 (11.3-12.8)\\
$N_{\rm H}(10^{21}{\rm cm}^{-2})$$^{\dagger}$ & 0.96 (0.72-1.2)       & 0.41 (fixed)    \\
normalization$^{\ddagger}$           & 8.64 (8.50-8.78)   & 2.19 (2.18-2.21)\\
$\chi^2/{\rm d.o.f.}$             & 66/72              & 55/73           \\ \hline
\end{tabular}
\end{center}
\vspace*{1em}\par\noindent\hspace*{4em}
$\ast $ The numbers in parenthesis represent the 90\% confidence range.
\par\noindent\hspace*{4em}
$\dagger$ The Galactic line-of-sight absorption value for the bremsstrahrung model.
\par\noindent\hspace*{4em}
$\ddagger $ Flux (in units of $10^{-12}$ erg ${\rm cm}^{-2}$ ${\rm s}^{-1}$)
over the 2--10 keV energy range for\\
\noindent\hspace*{5em}
the power-law model, and normalization ($\times 10^{-3}$) of the $bremss$ model in $xspec$ 10.0.
\end{table*}

%
%

\begin{table*}[b]
\begin{center}
Table 3. The results of the past observations of RX~J0957.9+6903.\\
\vspace*{1em}
\begin{tabular}{cccc}
\hline\hline
Instrument	   &	Date	  &	band [keV]	&	flux$^\ast$\\ \hline
Einstein IPC$^\dagger$&  1979 Apr 27 &     0.2--4	        & 	4.1 \\
Einstein HRI$^\dagger$&  1979 May\hspace*{.75em}3 &     0.2--4	        & 	3.5 \\
EXOSAT CMA$^\dagger$  &  1983-1986   &     0.05--2.0       &       1.0\\ 
ROSAT PSPC$^\ddagger$ &  1991 Mar 25 & 	0.5--2.0	&	2.0 $\pm$ 0.1\\
	           &  1991 Oct 16 & 	0.5--2.0	&	2.7 $\pm$ 0.1\\
	           &  1992 Sep 29 & 	0.5--2.0	&	1.7 $\pm$ 0.1\\	\hline
ASCA  GIS$^{\P}$     &  1993--1999  & 	0.2--4		&	6.3 $\pm$ 0.1\\
		   &              & 	0.05--2.0	&	3.3 $\pm$ 0.1\\
		   &  		  & 	0.5--2.0	&	3.1 $\pm$ 0.1\\ \hline
\end{tabular}
\end{center}
\vspace*{1em}\par\noindent\hspace*{4em}
$\ast $ In units of $10^{-12}$ erg ${\rm cm}^{-2}$ ${\rm s}^{-1}$.
\par\noindent\hspace*{4em}
$\dagger $ Einstein and EXOSAT results quoted from Fabbiano (1988), 
\par\noindent\hspace*{4.75em}
Giommo et al. 1991, respectively.
\par\noindent\hspace*{4em}
$\ddagger $ Assuming the power-law model with the free absorption and photon index.
\par\noindent\hspace*{4em}
${\P} $ Assuming the absorbed power-law model with the best fit parameters
\par\noindent\hspace*{4.75em}
in table 2.
\end{table*}

\bigskip

\onecolumn
\large

\vspace*{5em}
\begin{center}
\hspace*{1em}
\psfig{file=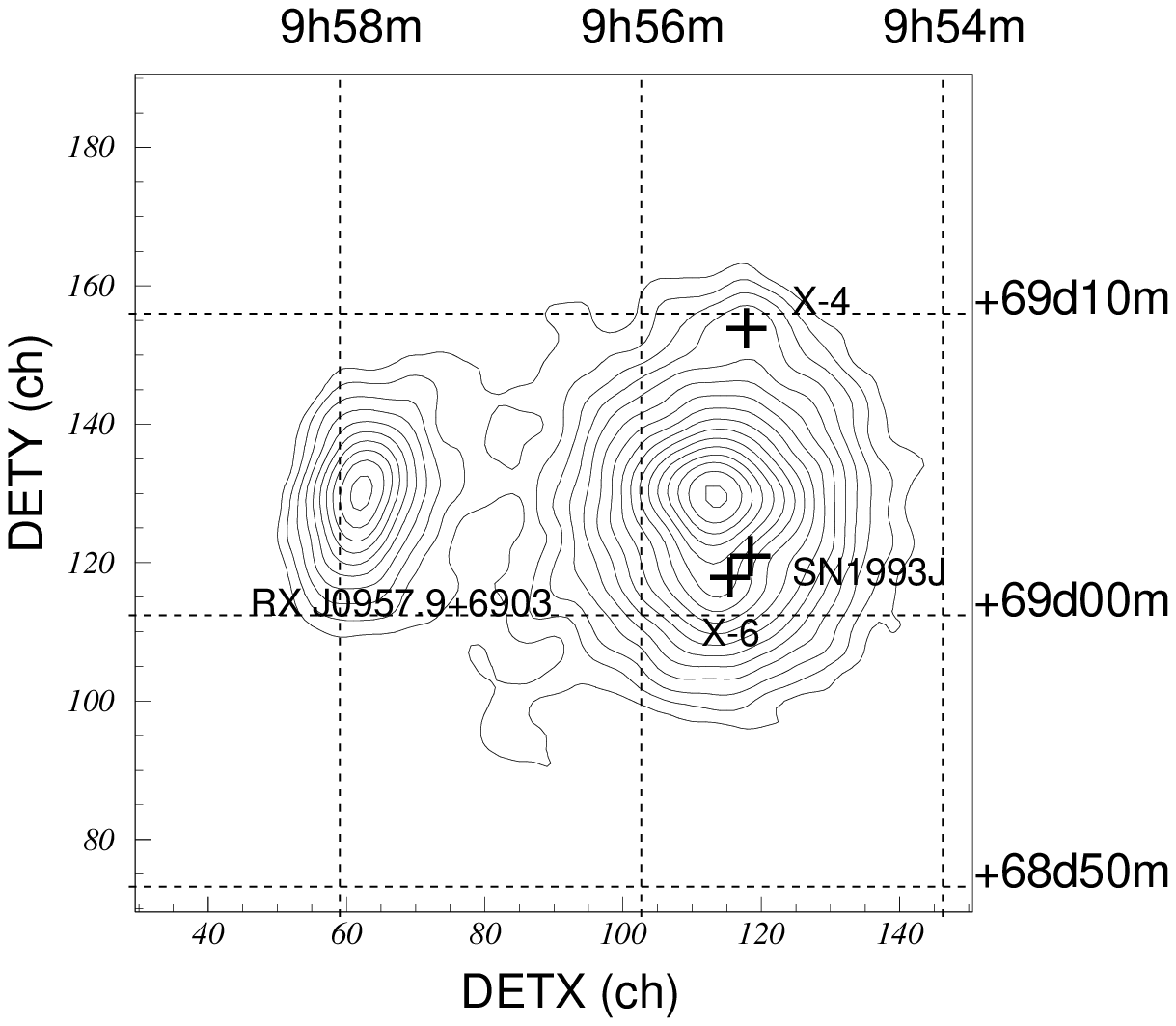,width=15cm}
\end{center}

\vspace*{5em}

{Fig. 1. The 0.7--10 keV X-ray image of the M81 region obtained by summing the
GIS2 + GIS3 data for the 7th - 9th observation and smoothing
with a Gaussian distribution of $\sigma={30}^{\prime\prime}$.
The background is included, and the image is not corrected for vignetting.
The contour level are logarithmic. Sky coordinates are J2000.
The crosses represent the X-ray location of SN1993J, X-6 and X-4.}

\clearpage

\begin{center}
\vspace*{-1em}
\hspace*{-0.1em}
\psfig{file=fig2a.ps,angle=-90,width=14cm}
\end{center}

\bigskip

\centerline{
\psfig{file=fig2b1.ps,angle=-90,width=8cm}
\hspace*{2em}
\psfig{file=fig2b2.ps,angle=-90,width=8cm}
}

\vspace*{2em}

{Fig 2. (a) Long-term light curves of RX~J0957.9+6903 binned
into each observation, 1st - 16th, obtained with GIS2+GIS3.
The error bars correspond to the Poisson error.
Note that the flux is determined from the spectrum of each observation
in terms of an absorbed power-law model. The photon index and absorption column
density are free for 
all the spectra, except the 16th observation 
in which the absorption was fixed at  the Galactic value (see text).\\

(b) Two examples of short-term light curves of RX~J0957.9+6903
with 1/4-day bin, obtained with GIS2+GIS3.
The error bars correspond to the Poisson error.
The left panel is in the longest 3rd observation.
The right panel is in the 13th observation 
when the short--term variability was maximum, 25\%.
Also the $\chi^2$ value and d.o.f.
calculated against the hypothesis of constant intensity
are shown in the top right of each panel.
The solid line shows the average of each observation.}

\clearpage

\centerline{
\hspace*{-1em}
\psfig{file=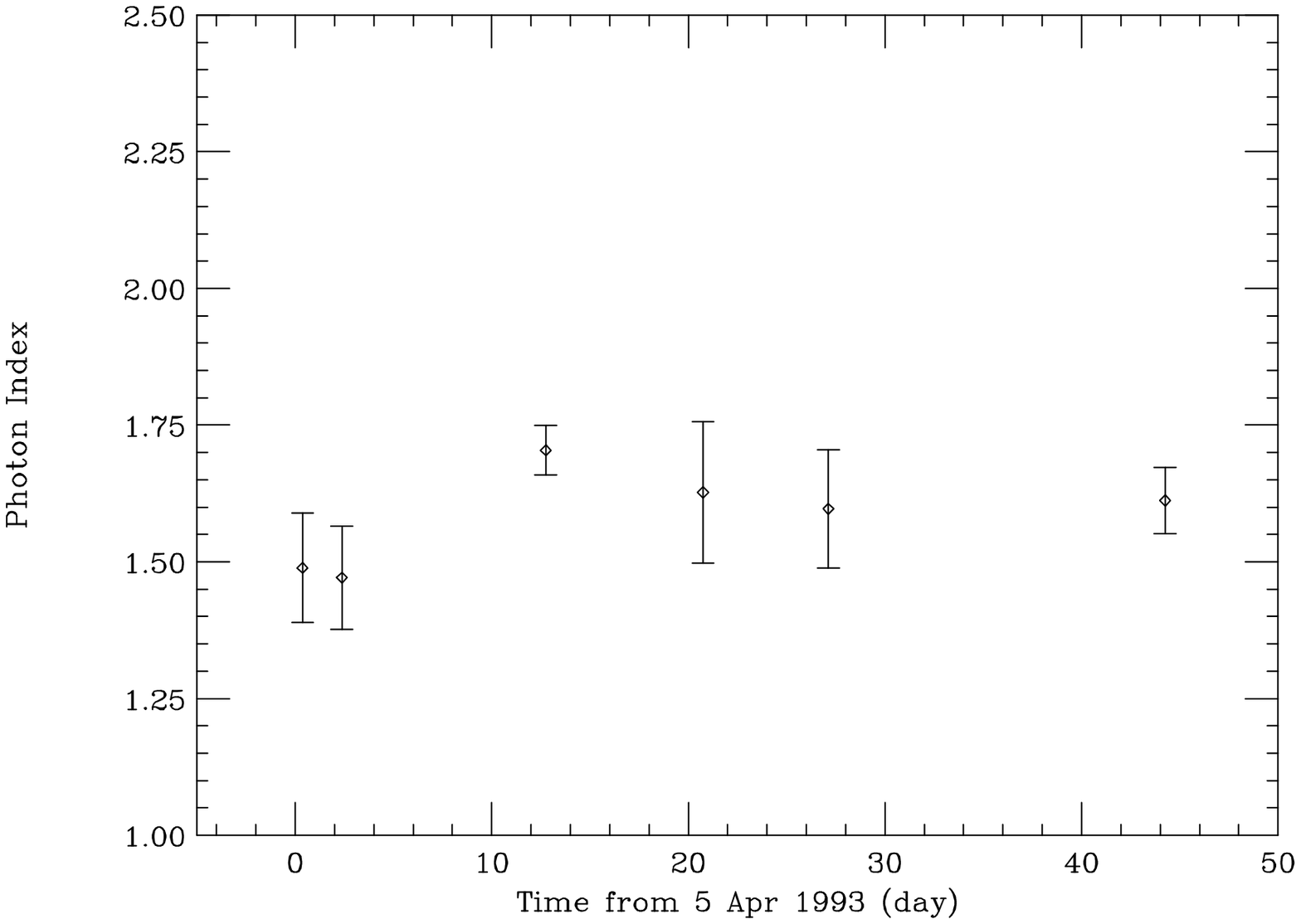,width=9.3cm}
\hspace*{1em}
\psfig{file=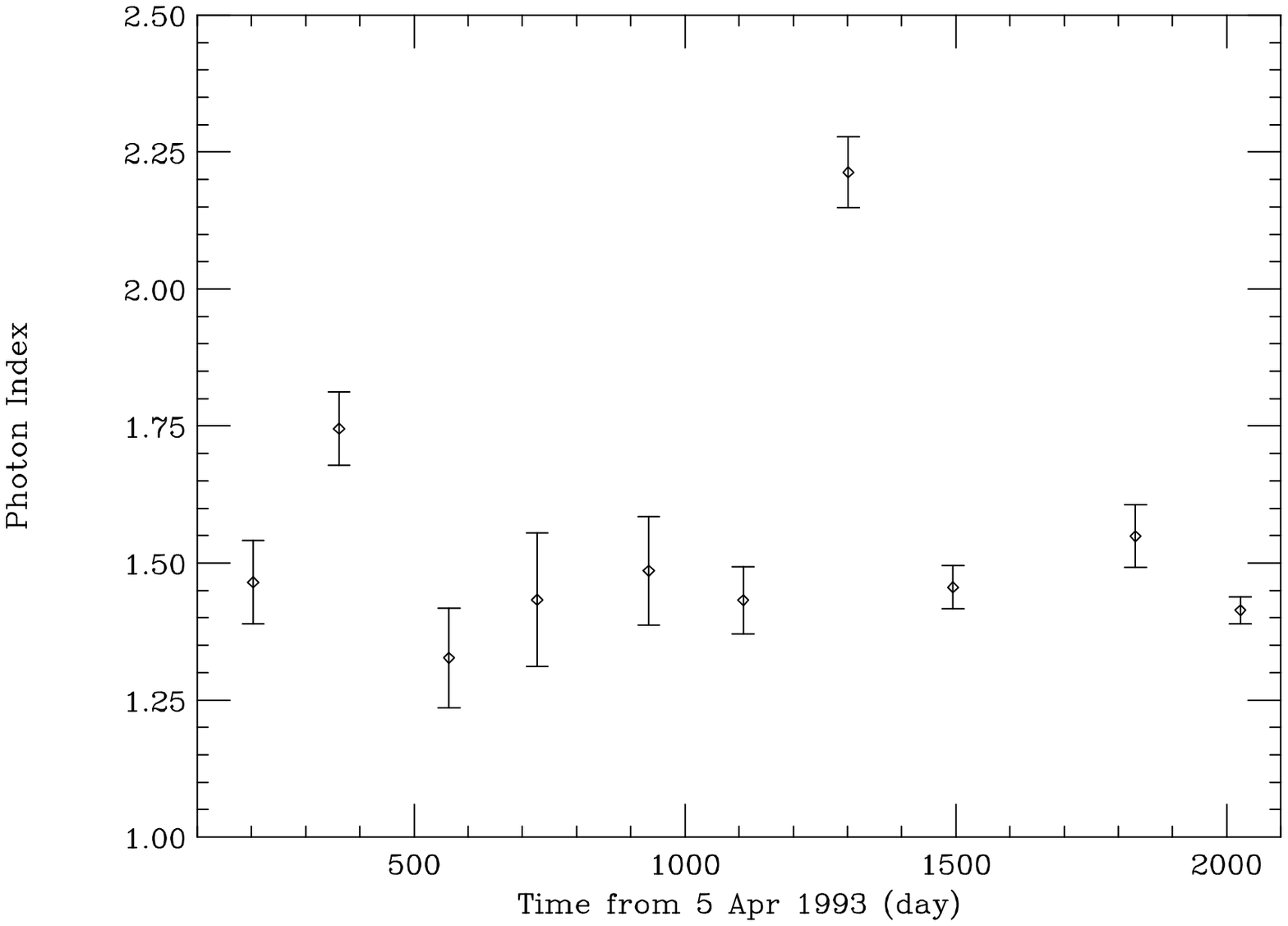,width=9.3cm}
}

\vspace*{3em}

{Fig 3. Variability of photon index $\Gamma$ with 90\% confidence errors.
The fitting model is an absorbed power-law.
Absorption and photon index are taken as free parameters
except 16th observation in which the absorption is fixed at the Galactic value.
The left and right panels show results from the
1st to 6th and 7th to 16th observations, respectively.
Note the difference in time scale.}

\vspace*{8em}

\centerline{
\psfig{file=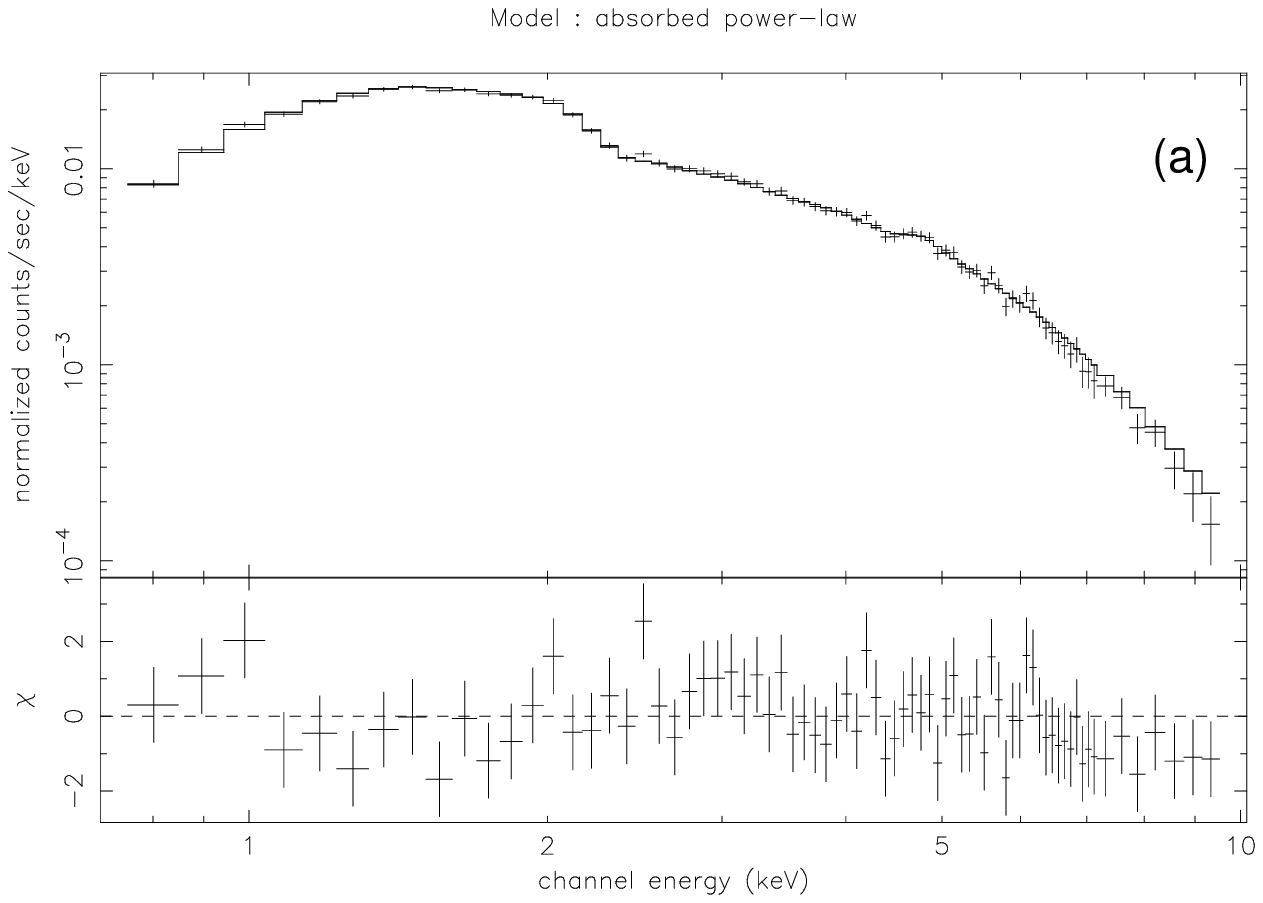,angle=-90,width=9cm}
\hspace*{2em}
\psfig{file=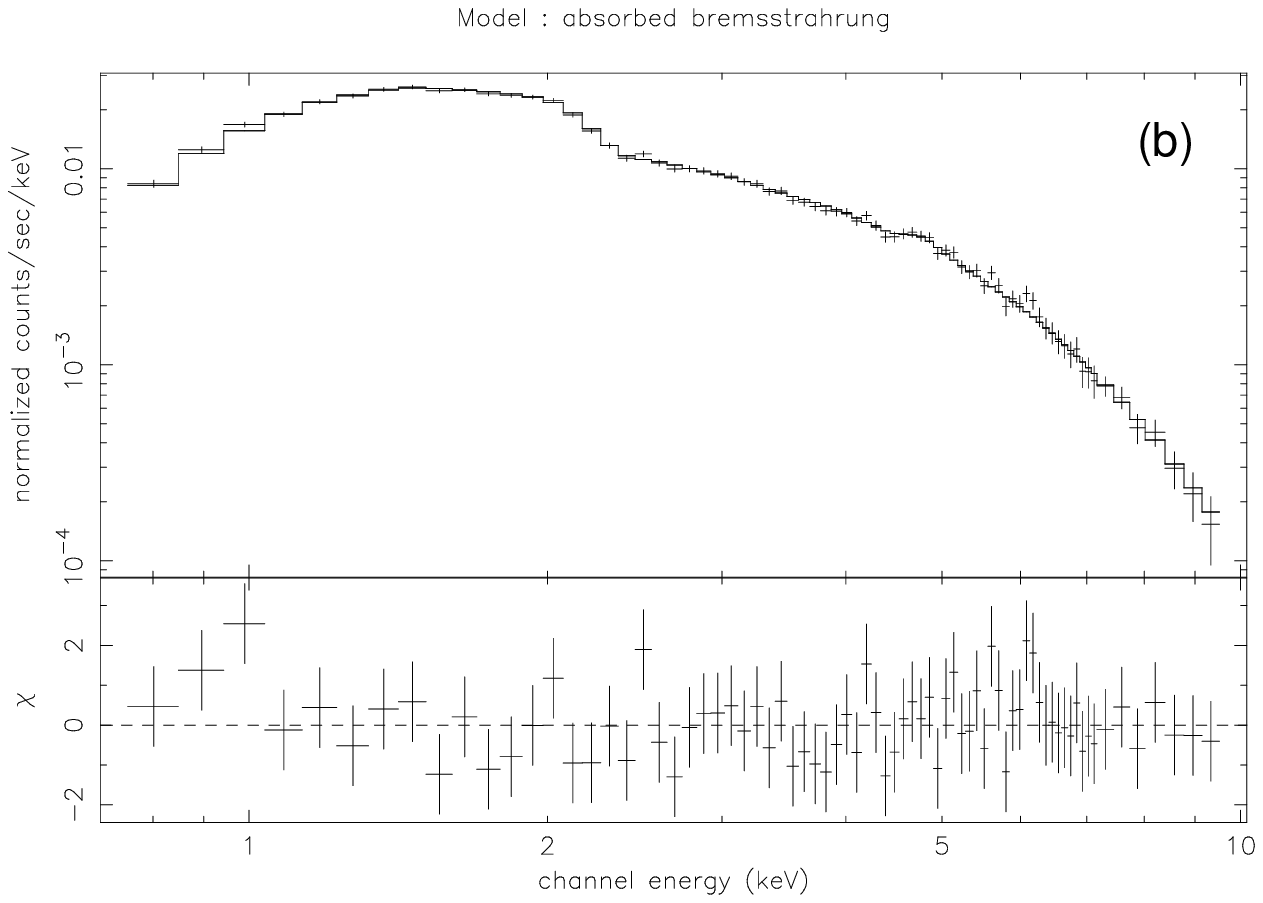,angle=-90,width=9cm}
}

\vspace*{3em}

{Fig 4. The GIS spectrum of RX~J0957.9+6903 in the 0.7-10 keV band,
summed over observations 1-16 except 13 and 16.
The solid line shows the best-fit power-law model (panel a) and
bremsstrahlung model (panel b) convolved through respective instrumental
responses. The fit residuals are shown in lower panels. }

\clearpage

\vspace*{5em}

\centerline{
\psfig{file=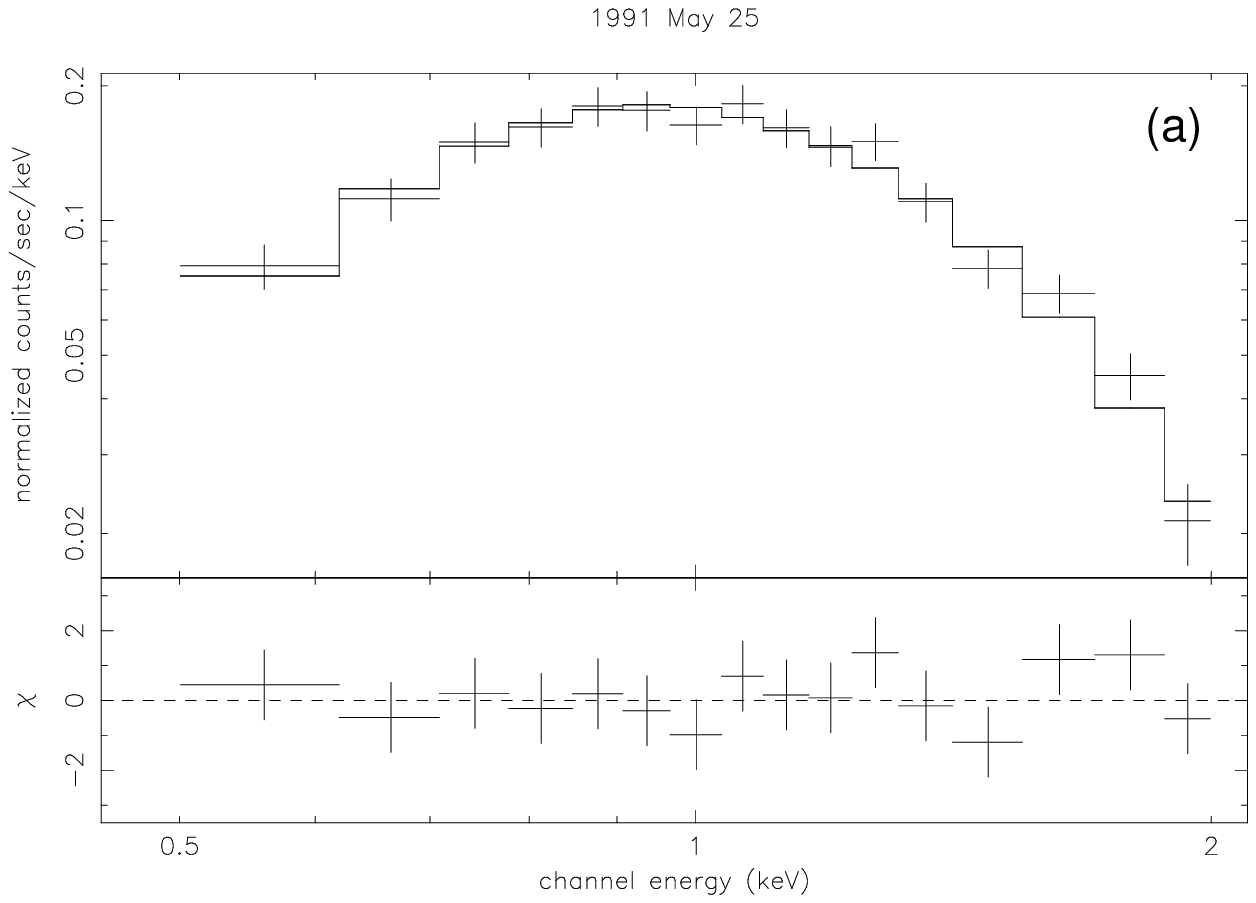,angle=-90,width=9cm}
\hspace*{2em}
\psfig{file=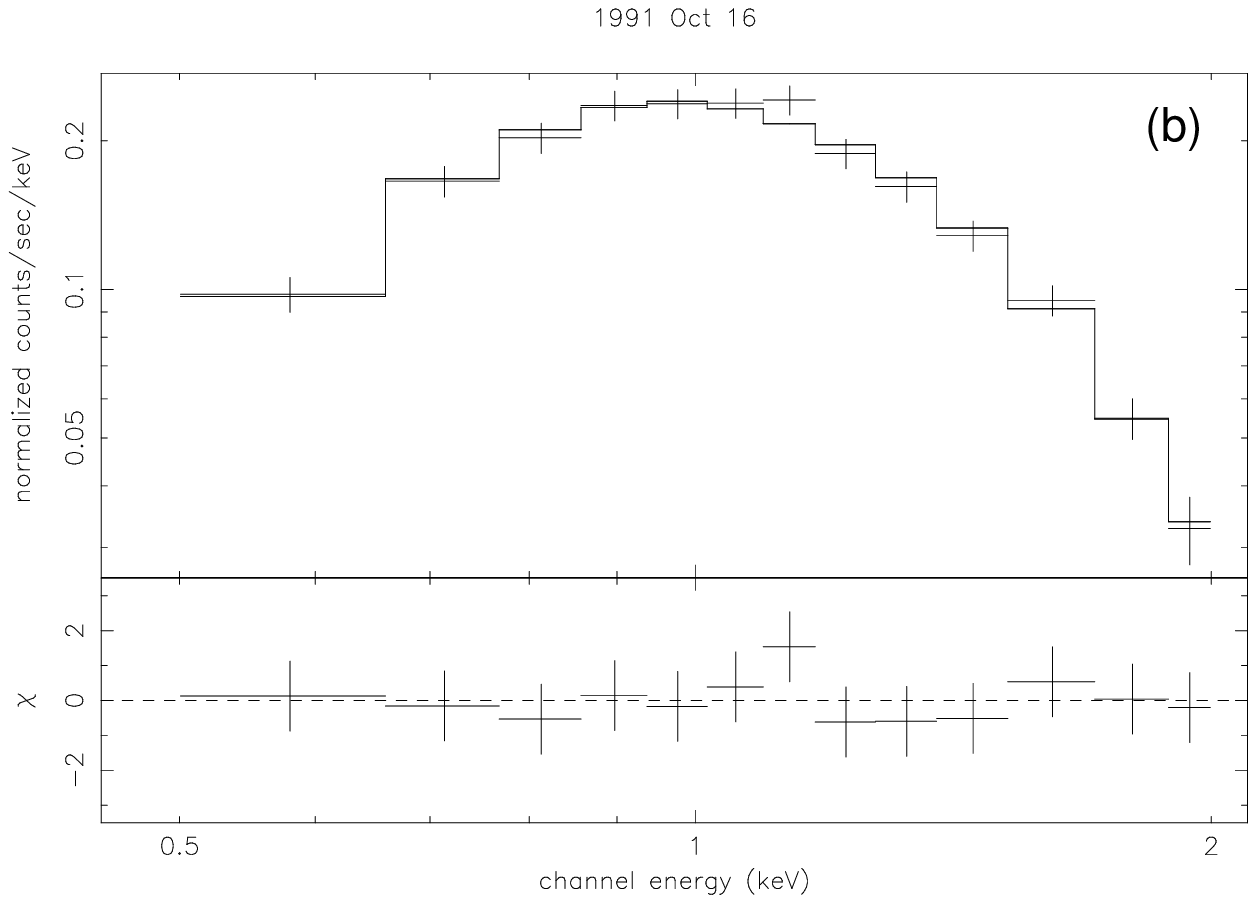,angle=-90,width=9cm}
}

\centerline{
\psfig{file=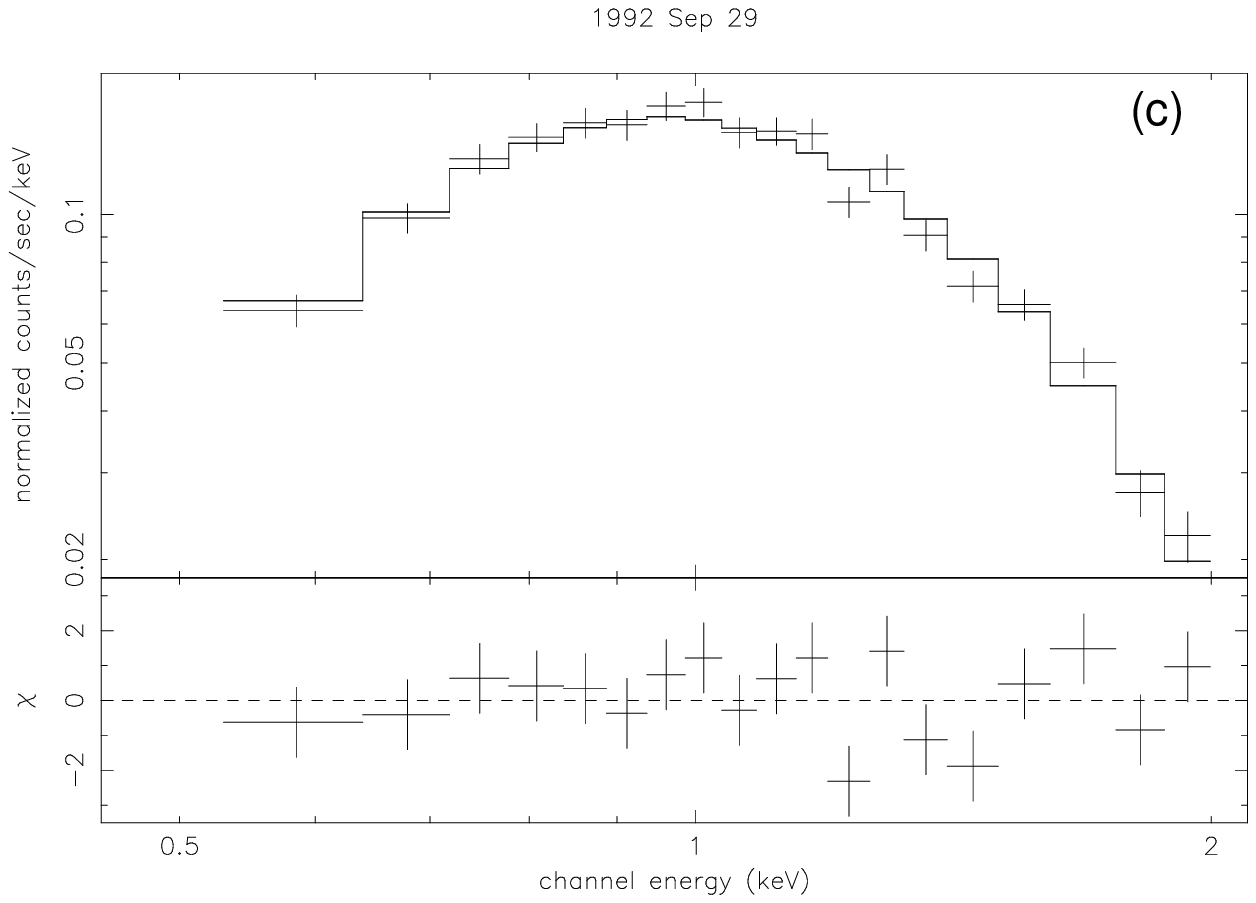,angle=-90,width=9cm}
}

\vspace*{5em}

{Fig 5. PSPC spectra of RX~J0957.9+6903 in the 0.5-2.0 keV band,
obtained on 1991 May 25 (panel a), October 16 (panel b) and 1992 September 29 (panel c).
The solid line shows the best-fit power-law model
convolved through the instrumental response.
The fit residuals are shown in lower panels. }

\clearpage

\vspace*{5em}
\centerline
{
\psfig{file=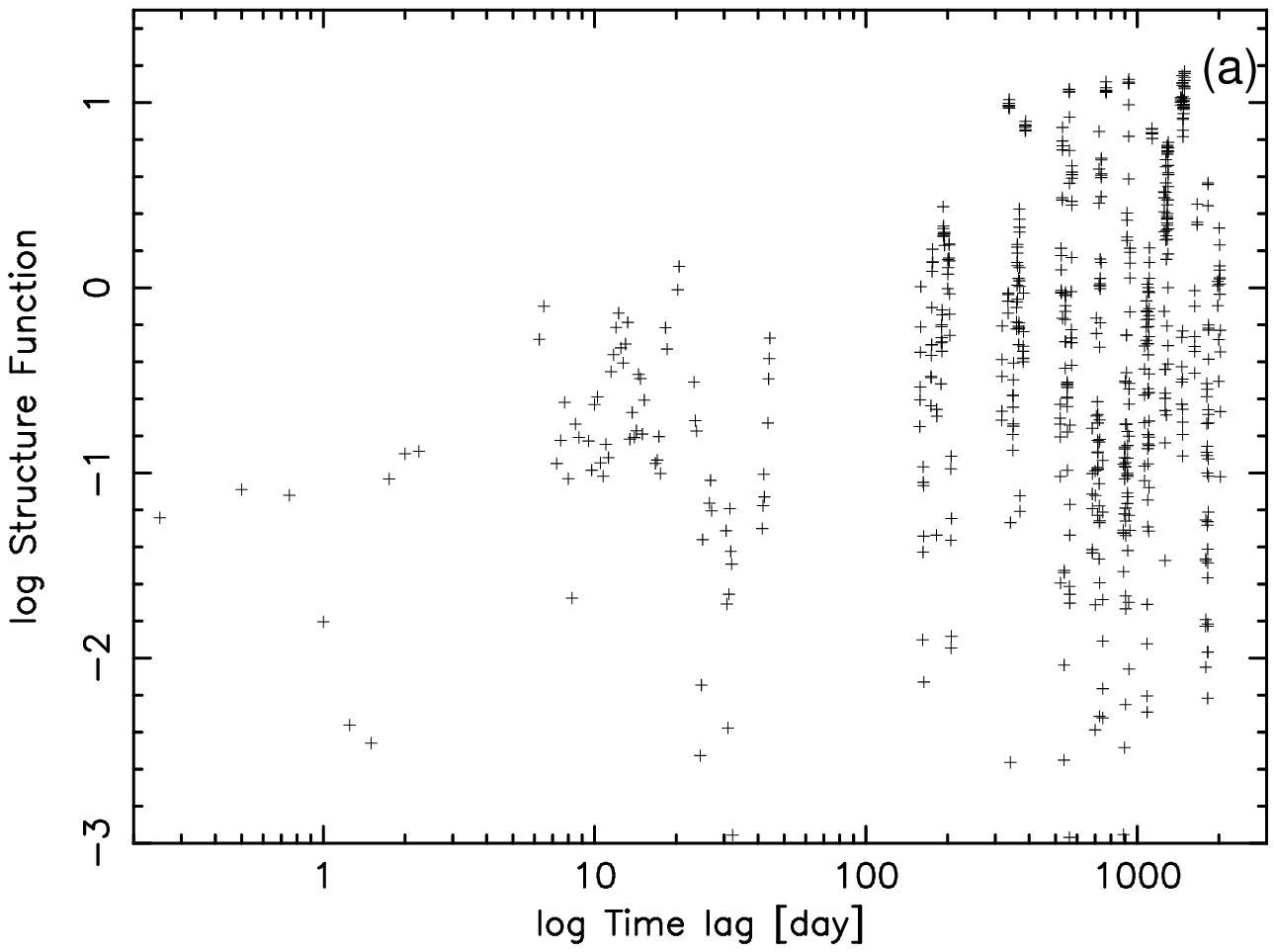,angle=-90,width=9cm}
\hspace*{1em}
\psfig{file=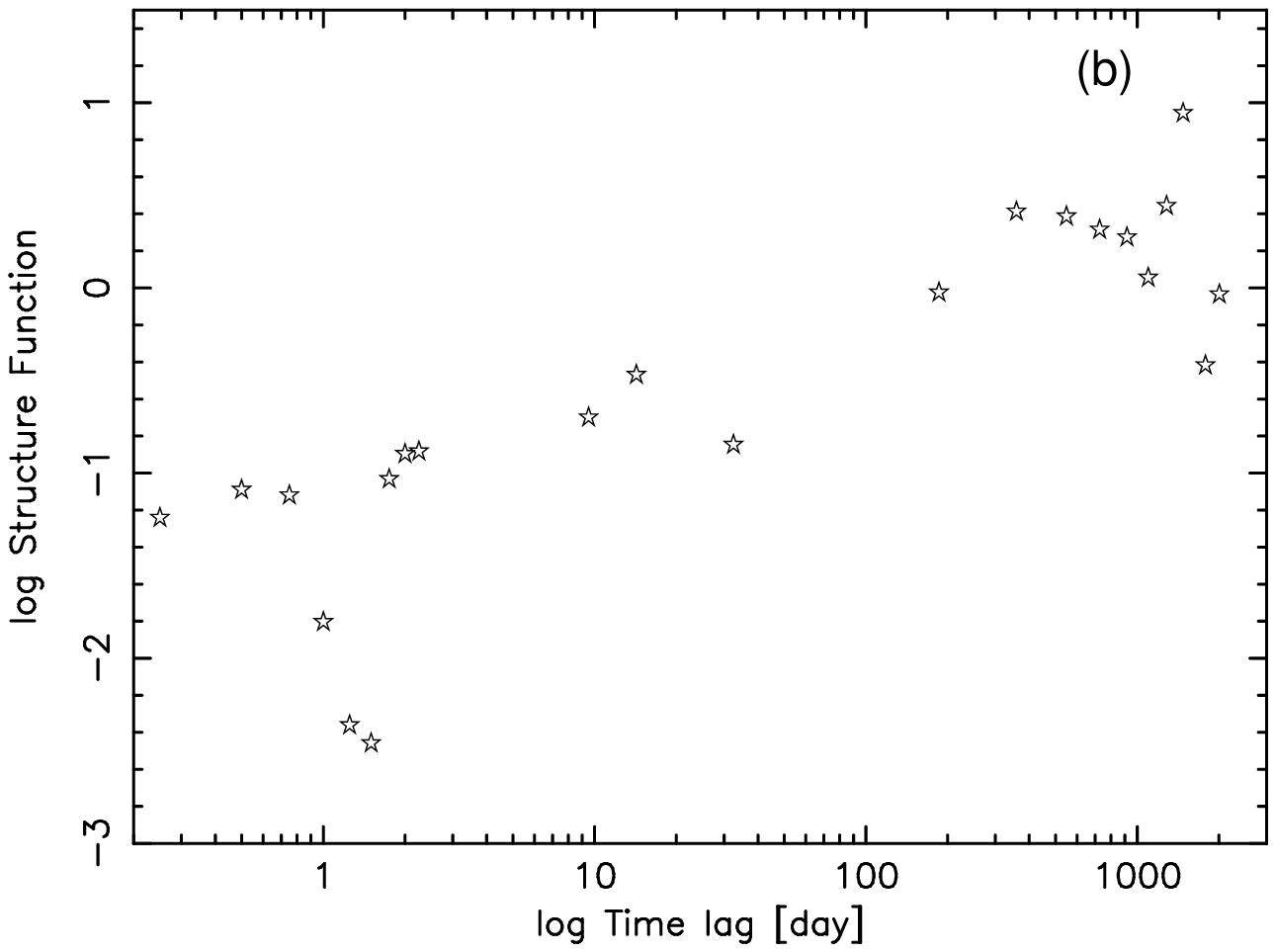,angle=-90,width=9cm}
}

\vspace*{5em}

\centerline
{
\psfig{file=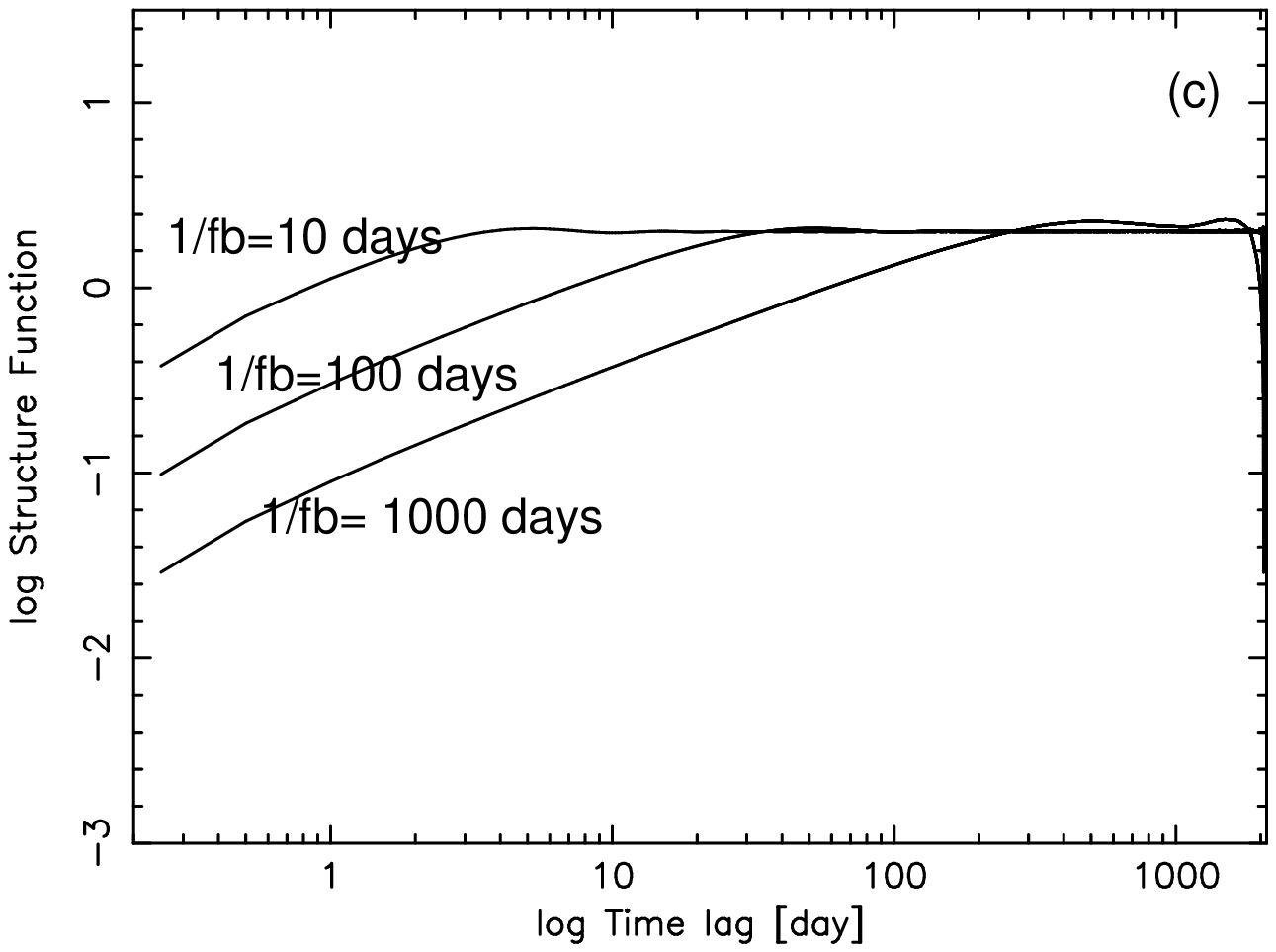,angle=-90,width=9cm}
\hspace*{1em}
\psfig{file=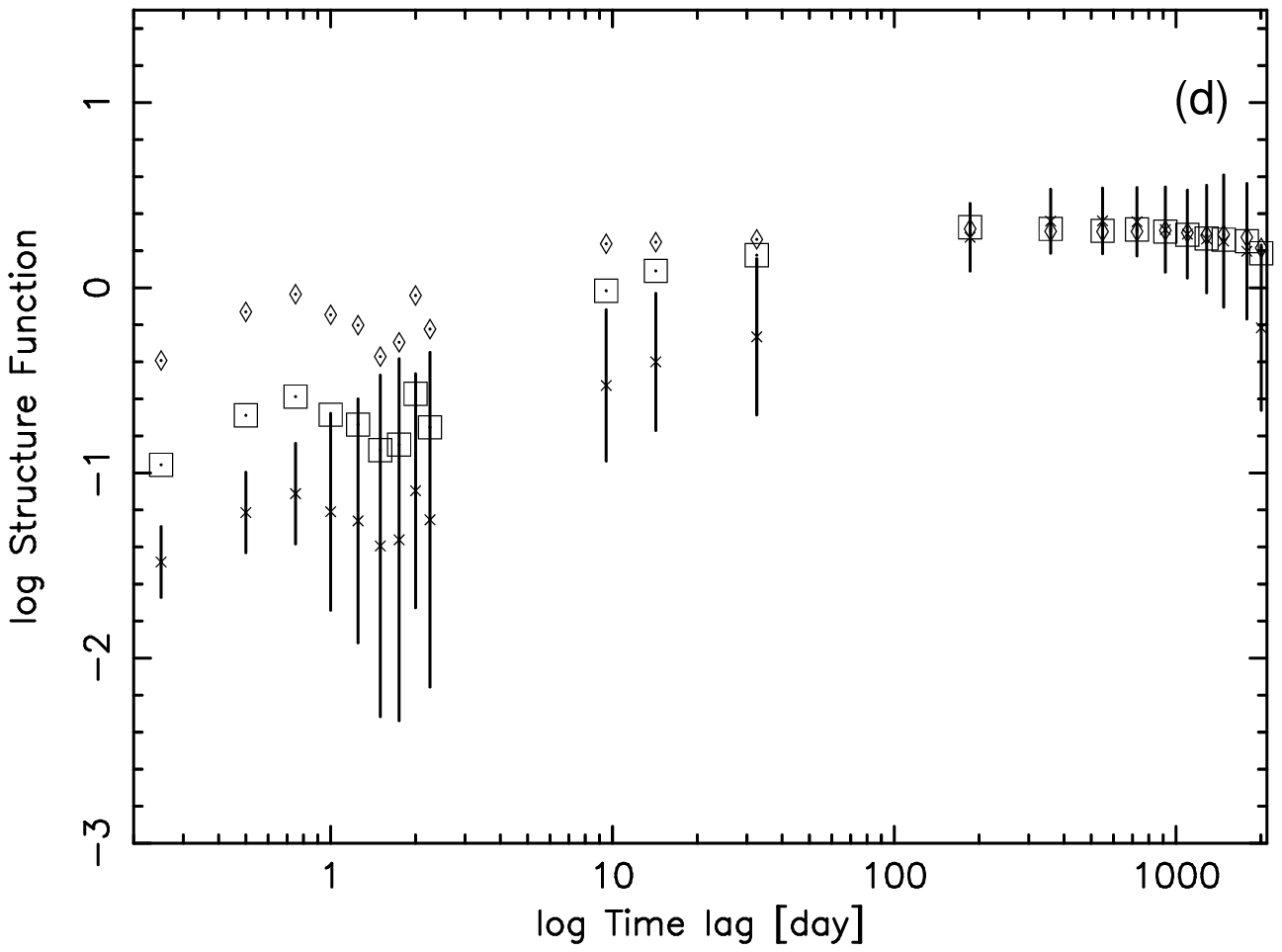,angle=-90,width=9cm}
}

\vspace*{5em}

{Fig 6. (a) SF of RX~J0957.9+6903 before binning. (b) SF after binning. 
(c) Ensemble average of 1000 fake SFs without window function, generated from the PSD having $\alpha = -1.55$ and $1/f_{\rm b}$ = 10, 100 and 1000 days.
(d) Ensemble average of 1000 fake SFs after applying the window function
and binning, generated from the same PSD as (c).
Diamond, square and cross shows $1/f_{\rm b}$ = 10, 100 and 1000 days,
respectively. Only for cross points, we display error bars,
which are dispersions among the 1000 simulations.}

\clearpage

\centerline
{
\psfig{file=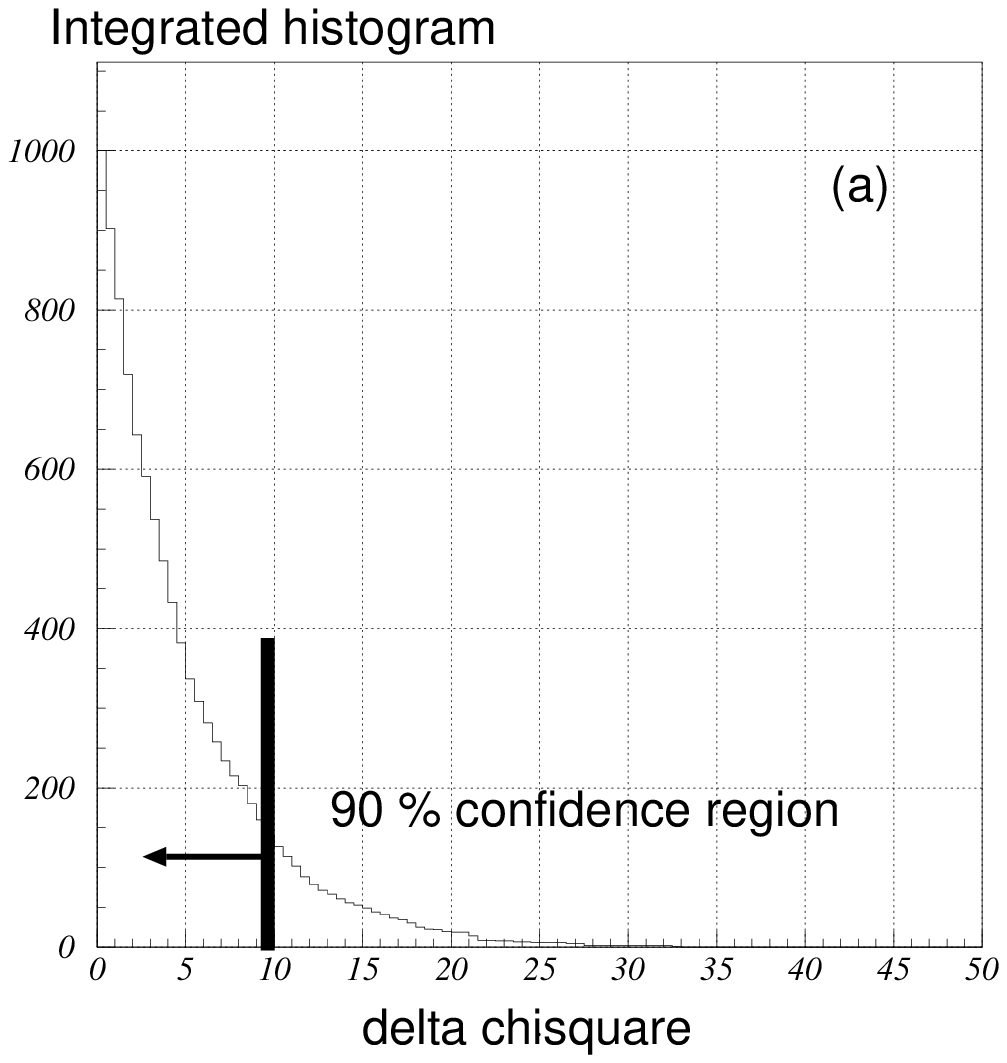,width=8.8cm}
\hspace*{1em}
\psfig{file=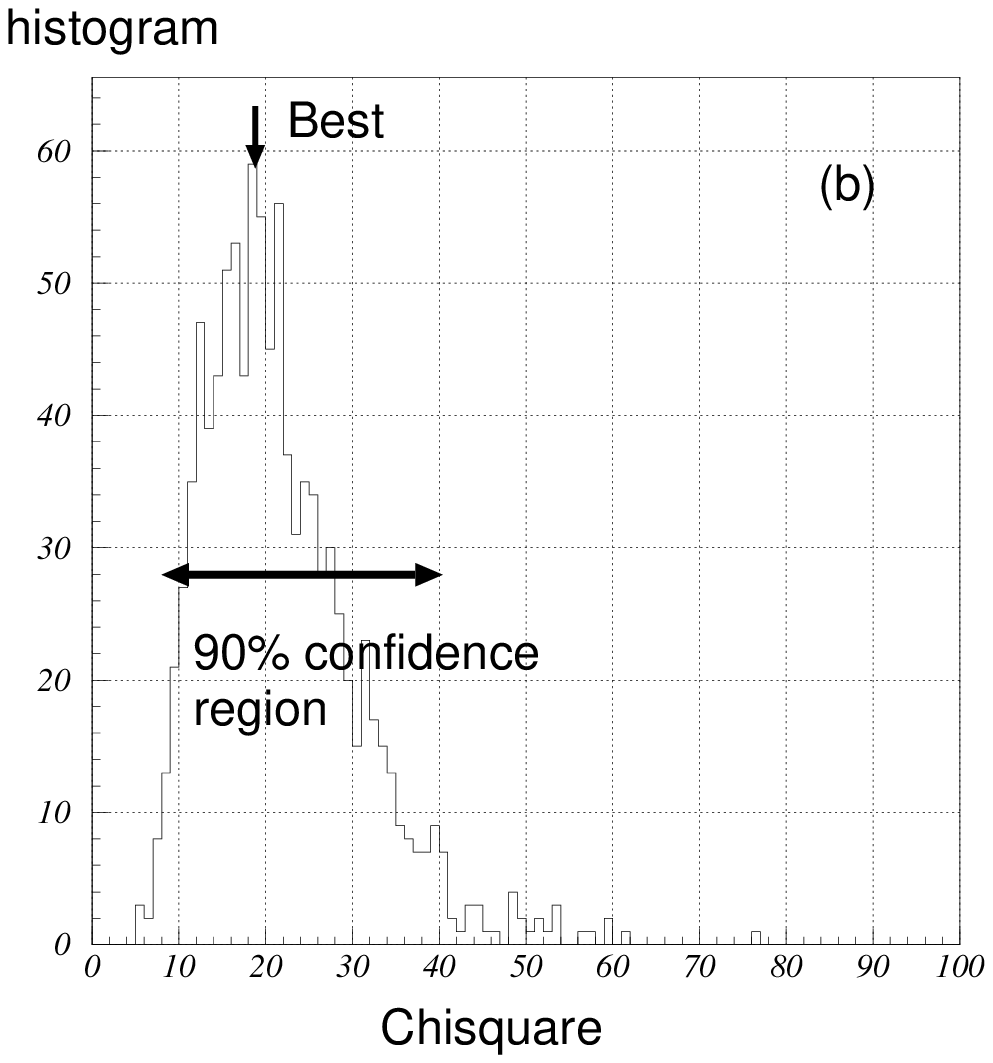,width=8cm}
}

\vspace*{3em}

{Fig 7. (a) The integrated distribution of $\Delta\chi^2 = {\chi^2}_{\rm true}
- {\chi^2}_{\rm min}$ of 1000 simulated light curves for
$1/f_{\rm b} = 10^{3.2}$ day and $\alpha = -1.55$.
(b) The $\chi^2$ distribution of 1000 fake light curves (histogram) and
the $\chi^2$ at the best parameters of $f_{\rm b}$ and $\alpha$ (arrow).
Note that degree of freedom of the binned SF is 22.}

\vspace*{3em}

\centerline{
\psfig{file=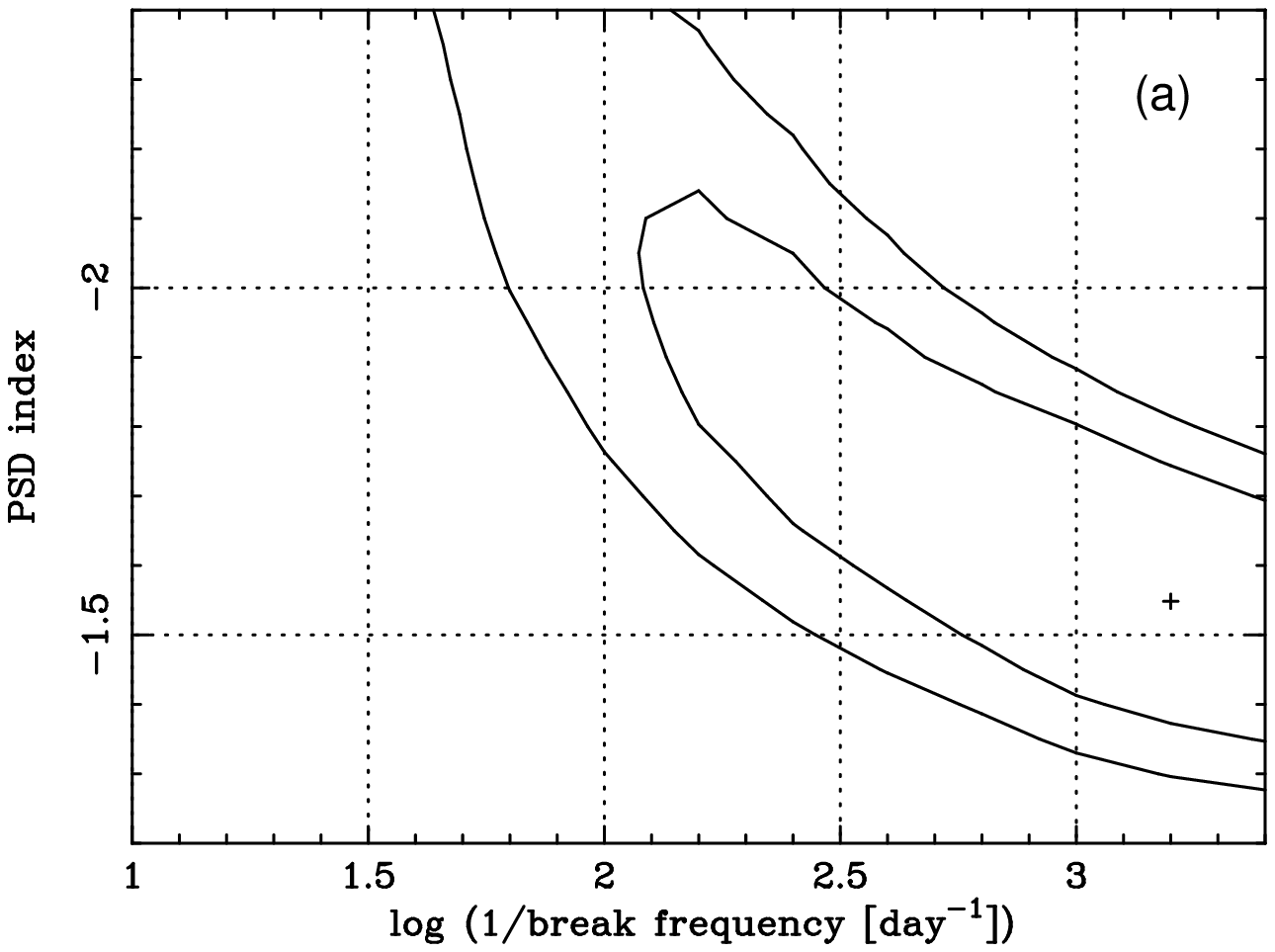,angle=-90,height=6.6cm}
\hspace*{1em}
\vspace*{-6.63cm}
\psfig{file=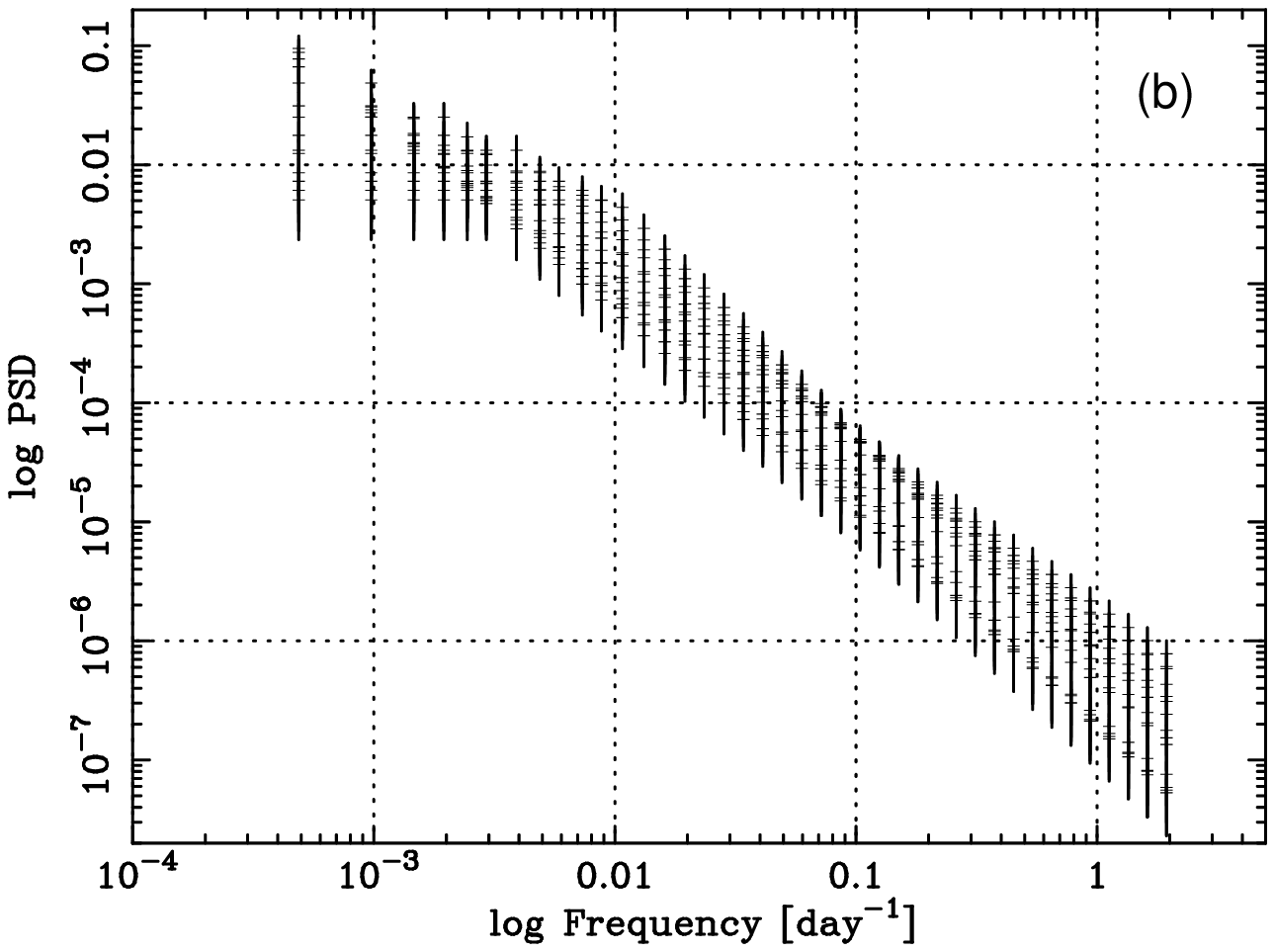,angle=-90,height=6.6cm}
}

\vspace*{20em}

{Fig 8. (a) 90\% and 99\% confidence regions in the 2-dimension plane of
$f_{\rm b}$ and $\alpha$. The cross represents the best fit parameters from
the 1/4-day binned light curve of RX~J0957.9+6903.
(b) Estimated PSD corresponding to the 90\% confidence region of panel (a). }

\end{document}